\documentclass[12pt,preprint]{aastex}

\newcommand{\myemail}{lhs@usyd.edu.au}
\def\g296{G296.5+10.0}

\def\ga{\ifmmode\stackrel{>}{_{\sim}}\else$\stackrel{>}{_{\sim}}$\fi} 
\newcommand\HI{H\,{\sc i}}

\shorttitle{Faraday rotation of an SNR: A magnetized progenitor wind?}
\shortauthors{Harvey-Smith et al.}

\begin{document}

\title{Faraday rotation of the supernova remnant G296.5+10.0: Evidence for a Magnetized Progenitor Wind}

\author{L. Harvey-Smith,\altaffilmark{1} B. M. Gaensler,\altaffilmark{1,5} 
R. Kothes,\altaffilmark{2} R. Townsend,\altaffilmark{3}
G. H. Heald,\altaffilmark{4} C.-Y. Ng\altaffilmark{1} and A. J.
Green\altaffilmark{1}}
\altaffiltext{1}{Sydney Institute for Astronomy (SIfA), School of Physics, 
The University of Sydney, NSW 2006, Australia; \myemail}
\altaffiltext{2}{National Research Council of Canada, Herzberg
Institute of Astrophysics, Dominion Radio 
Astrophysical Observatory, Penticton, British Columbia, V2A 6J9, Canada}
\altaffiltext{3}{Department of Astronomy, University of Wisconsin, 
Madison, WI 53706}
\altaffiltext{4}{ASTRON, 7990 AA Dwingeloo, The Netherlands}
\altaffiltext{5}{Federation Fellow, Australian Research Council}

\begin{abstract}

We present spectropolarimetric radio images of the supernova remnant (SNR) \g296 at frequencies near 1.4~GHz, observed with the Australia Telescope Compact Array. By applying rotation measure (RM) synthesis to the data, a pixel-by-pixel map of Faraday rotation has been produced for the entire remnant.  We find \g296 to have a highly ordered RM structure, with mainly positive RMs (mean RM of $+$28 rad~m$^{-2}$) on the eastern side and negative RMs (mean RM of $-$14 rad~m$^{-2}$) on the western side, indicating a magnetic field which is directed away from us on one side and toward us on the other. We consider several possible mechanisms for creating the observed RM pattern. Neither Faraday rotation in foreground interstellar gas nor in a homogeneous ambient medium swept up by the SNR shell can easily explain the magnitude and sign of the observed RM pattern. Instead, we propose that the observed RMs are the imprint of an azimuthal magnetic field in the stellar wind of the progenitor star. Specifically, we calculate that a swept-up magnetized wind from a red supergiant can produce RMs of the observed magnitude, while the azimuthal pattern of the magnetic field at large distances from the star naturally produces the anti-symmetric RM pattern observed. Expansion into such a wind can possibly also account for the striking bilateral symmetry of the SNR's radio and X-ray morphologies.

\end{abstract}


\section{Introduction}

Observations of radio supernova remnants (SNRs) provide a unique probe of the circumstellar environments of supernova progenitor stars and the interstellar medium (ISM). In particular, the interactions between SNRs and their environments give rise to a wide range of observable effects from which we can derive important physical insights. Spectral line maps of neutral hydrogen ({\sc Hi}) in the vicinity of SNRs enable us to calculate the age, swept-up shell mass and energy of the supernova explosion \citep{gmg98,vdg+02,frk04}. Combined with model rotation curves for the Galaxy, {\sc Hi} data may be used to constrain the kinematic distance to the remnant \citep{plm+93,tl08}. Under certain conditions, regions of interaction between a SNR shock and a dense molecular cloud produce hydroxyl masers, allowing us to derive the properties of the SNR shock, estimate the distance to the remnant and to calculate the postshock magnetic field through the Zeeman effect \citep{fgs94,cfg+97}. In many SNRs there is evidence of physical confinement by denser, cooler surrounding material, which allows us to probe the physical, chemical and magnetic properties of molecular clouds \citep{bgbw88,fgs94,gfgo97,che99} and provide information on particle acceleration and the production of cosmic rays \citep[e.g.,][]{kpg+95,thk+98,fmt+03}.

The appearance of young SNRs can be strongly influenced by the mass-loss history of the progenitor star \citep{che82c,ftb+91,kh95}. For example, the properties of the progenitor stellar wind have been inferred in the young SNR Cassiopeia A by studying the dynamics of the clumpy shock boundary in the remnant and calculating the swept-up shell mass \citep{co03,vkb98}. In addition, the structure and brightness evolution of the radio remnant of supernova 1987A has been shown to have been shaped by successive interactions of the remnant with material from stellar winds originating from the blue- and red-supergiant phases of the progenitor star \citep{cf87,gms+97}.  Interaction of the supernova ejecta with stellar winds is a phenomenon that occurs in the earlier or mid-stages of a radio SNR's life \citep{dwa05} although in some cases the passage of the remnant through the circumstellar material (CSM) can continue for many thousands of years in large stellar wind bubbles \citep{lrr+99,gva06}.

Magnetic fields are known to influence the development of supernova remnants \citep{kbk+65,che74}.  This can be as a direct consequence of magnetic confinement of the expanding SNR shell under the influence of external ISM magnetic fields \citep{laa62a,wg68}, or due to the sweeping-up of a CSM whose distribution is already anisotropic due to magnetic channeling and confinement of the pre-supernova wind(s) \citep[see, e.g.,][]{uo02,uot08}.

Alternatives to the direct shaping of SNRs by magnetic confinement have also been proposed. \cite{bls90} asserted that the eventual shape of a supernova remnant is primarily influenced by the shape of the cavity swept out by the progenitor star. In some cases, the shape of such a stellar wind bubble will be dominated by magnetic fields. In a theoretical treatment of this problem, \cite{cl94} considered the case of a bubble driven by a magnetized stellar wind. They found that, in a bubble resulting from a stellar wind with a toroidal magnetic field, the magnetic tension in the equatorial region of the shell can confine the bubble and cause elongation in the polar direction. The proposal that SNRs can be shaped by a  magnetically-influenced cavity was arrived at by \cite{gae98}, who proposed that SNRs with barrel morphologies (remnants with negligible emission along the axis of symmetry and bright limbs of emission perpendicular to this axis) form in ISM bubbles that are elongated by the Galactic magnetic field. Whether this elongation is due directly to magnetic confinement or the product of density stratifications along the local magnetic field axis is less clear.  

All of these different mechanisms will have an effect at some point in the life of a SNR. It is likely that very young remnants are dominated by the supernova explosion itself, middle-aged remnants will primarily be sculpted by the CSM and older remnants will be shaped by the surrounding ISM.  Disentangling these different effects is a complicated task, but if the production and evolution of supernova remnants is to be understood, quantitative studies of the magnetic fields in their immediate environments are crucial.

A useful experimental probe of cosmic magnetic fields is Faraday rotation. When polarized electromagnetic radiation propagates through a magnetized ionized medium, its electric field vector is rotated through an angle which depends on the square of the wavelength and the path integral of the electron-weighted line-of-sight magnetic field strength. This effect is called Faraday rotation. The angle of rotation, $\Delta\chi$, is related to the square of the wavelength, $\lambda$, via $\Delta\chi = RM \lambda^2$ where the RM is the rotation measure and is defined as

\begin{equation}
\label{eq1}
{\rm RM} = 0.81 \int_{source}^{observer} n_e  B_{||} ~ \textit{dl}  ~{\rm rad~m^{-2}}
\end{equation}

where $B_{||}$ is the line-of-sight component of the magnetic field strength in $\mu$G, $n_e$ is the electron density in cm$^{-3}$ and $dl$ is measured in parsecs.  

Evaluating the RM of linearly polarized radio emission gives us an indication of the line-of-sight magnetic field orientation in the foreground Faraday rotating medium. Measurements of other physical cues, such as the interstellar rotation measure along similar lines of sight, can assist us in disentangling the Faraday rotation due to the Galactic foreground from the RM occuring internally to the source being studied. If the RM is intrinsic to the source, by measuring RMs at many points in an image we can learn about the line-of-sight orientation of the smooth component of the magnetic field in that source. In a supernova remnant, this will either tell us about the ambient magnetic field in the ISM, or the magnetic field in the swept-up shell of circumstellar material depending on the age and history of the remnant. 

In this paper we consider the Faraday rotation against \objectname{G296.5+10.0} (also known as PKS~1209--51/52), a high-Galactic latitude supernova remnant with a barrel morphology at both radio and X-ray wavelengths \citep{wg68,rmk+88,hb84}. The limbs of the remnant are believed to mark the material swept up by the blast wave from a core-collapse supernova as evidenced by the radio-quiet X-ray emitting neutron star (1E~1207.4--5209) at its center \citep{hb84,zpst00,zps04}. \g296\ displays a high degree of reflection symmetry \citep{ssmk92}, and lies 10 degrees above the Galactic plane at an estimated distance of 2.1 kpc \citep{gdg+00}.

Attempts to measure the polarization and rotation measure in \g296\ have previously been made by \cite{wg68}; \cite{dm76} and \cite{mh94b}, but these were limited in their accuracy because the RMs were derived by a linear fit to only two or three points in a plot of position angle vs. $\lambda^{2}$. This method of determining RMs introduces significant uncertainties into the quoted values, because the polarization position angle is known only as a multiplicative factor of $n\pi$ radians.  By using rotation measure synthesis \citep{bd05} to produce a well-sampled rotation measure spectrum at each image pixel, it is possible to eliminate these ambiguities and determine accurate RMs for the entire region. Another advantage of RM synthesis is that it allows us to resolve multiple RM components within a single line of sight. With the flexible correlator on the Australia Telescope Compact Array (ATCA) this can be achieved within a single 128~MHz frequency band.  

In this paper, we present full polarization and rotation measure synthesis images of \g296 using RM synthesis of ATCA data. In \S\ref{obs} we introduce the observations and methods of data analysis and in \S\ref{results} we show total intensity, polarization and rotation measure images and plots of RM distribution in \g296. The RM distribution and magnetic fields in \g296\ are discussed in \S\ref{discussion}. \S\ref{conclusions} describes our conclusions and suggestions for future work in this field.

\section{Observations \& Analysis}
\label{obs}
The supernova remnant \g296\ was observed between the 8th and the 11th of October 1998 using the ATCA. Analysis of the {\sc Hi} data from these observations was published by \cite{gdg+00}. We obtained the data from the ATCA archive in order to extract and analyze the (as yet unpublished) commensally recorded multi-channel continuum polarization information. The observations were carried out in the 210~m configuration, resulting in an angular resolution of 1$\farcm$5 at the central frequency of 1384~MHz. The total bandwidth of 128 MHz was correlated in 32~$\times$~4~MHz overlapping spectral channels. Those channels affected by self-interference, edge channels and overlapping channels were discarded to leave a useable bandwidth of 104~MHz split into 13~$\times$~8~MHz channels. Calibration terms were determined using observations of the sources PKS 1215-457 (gain solutions) and PKS 1934-638 (flux, polarization leakage and bandpass solutions) using the software reduction package {\sc miriad} \citep{sk03}. 109 individual pointings were combined to produce mosaiced images of the entire remnant in Stokes-\textit{I,Q,U,V} and polarized intensity. Images were deconvolved using the maximum entropy method using the {\sc miriad} task {\sc pmosmem} \citep{sbd99}. The RMS noise levels in the final continuum images were 2.0, 1.1 and 1.2 mJy~beam$^{-1}$ for Stokes-\textit{I,Q} and \textit{U} respectively. In order to make RM synthesis images, we also made separate mosaic images of each individual spectral channel for Stokes-\textit{Q} and \textit{U}, which had mean RMS noise levels of 3.4 and 3.5 mJy~beam$^{-1}$, respectively.  The maximum angular scale to which an interferometer is sensitive is set by the shortest spacing between antennas, in this case 31 metres. As a result, any emission greater than 24$^\prime$ is not detected in our ATCA observations, resulting in a Stokes-\textit{I} flux on large scales below the true value. Polarization images of the region are likely to contain almost all the flux however, as differential Faraday rotation in the Galactic foreground is expected to break the polarized image into smaller-scale structures.

Rotation measure synthesis \citep{bdb05} was employed to produce a full RM spectrum for each image pixel in steps of 20~rad~m$^{-2}$ between $-$2000 and $+$2000 rad~m$^{-2}$. The RM function had a FWHM of 530~rad~m$^{-2}$. We then used the spectral deconvolution algorithm {\sc rmclean} \citep{hbe09} with a maximum of 1000 iterations and a {\sc clean} cutoff of 0.0001 to remove the sidelobe pattern resulting from incomplete $\lambda^2$ sampling.  The outcome of these procedures was a spectrum of polarized flux as a function of RM at each pixel, with typical signal-to-noise ratio of $\sim$30 and a median RM uncertainty of 5.5~rad~m$^{-2}$.  Finally, an image of polarized intensity across the remnant was generated by using a 3-point polynomial fit around the peak of the deconvolved rotation measure spectrum at each pixel. Although we are able to confidently detect a polarized signal at a signal-to-noise ratio of $\sim4-5$, a meaningful RM can only be extracted for a somewhat stronger signal \citep{bdb05}. Accordingly, regions of the final polarized image with a signal below 10 $\times$ the root mean squared noise level in that image were masked (excluded) from the final maps of RM and polarization. At these high thresholds, polarization bias is negligible and so no debiasing was applied to any of the data.

\section{Results}
\label{results}
\begin{figure}
\includegraphics[angle=-90, width=140mm]{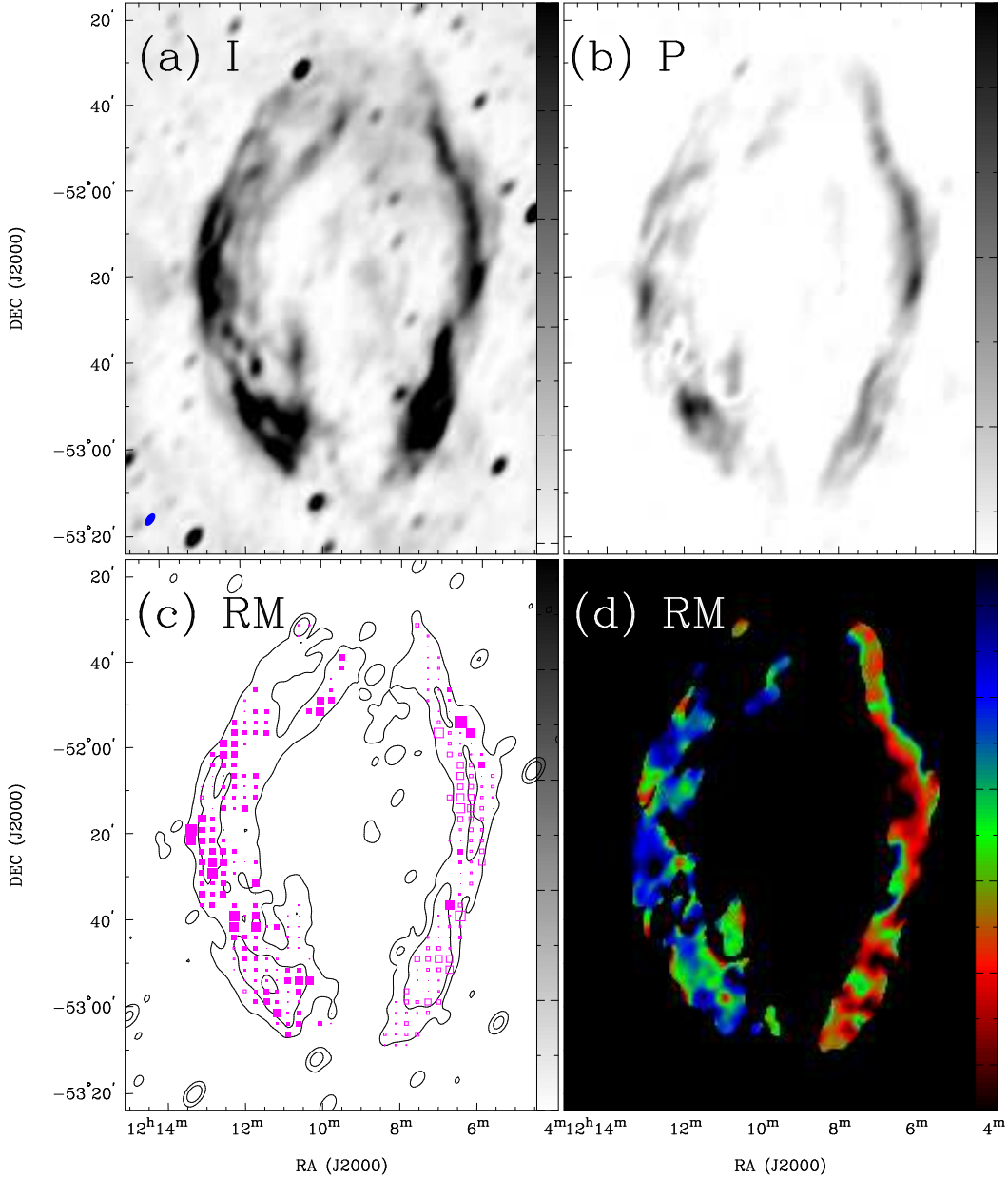}
\caption{ATCA images of SNR~\g296\ at 1.4~GHz. (a) Total intensity, shown with a greyscale ranging between $-$2 and $+$100 mJy~beam$^{-1}$.  The blue ellipse at the lower left is the angular resolution of the image, 3${\farcm}$3 $\times$ 1${\farcm}$8, at a position angle of $-$142$^{\circ}$ north through east; (b) Linearly polarized intensity, with greyscale ranging between 0.5 and 70 mJy~beam$^{-1}$; (c) Rotation measure plotted every five pixels (150$\arcsec$), with filled boxes representing positive RMs and unfilled boxes representing negative RMs. The linear dimensions of each box indicate the magnitude of the rotation measure. The range of box sizes corresponds to |RM| ranging from 0 to 128~rad~m$^{-2}$. Contours are the Stokes-\textit{I} image at $+$20 and $+$80 mJy~beam$^{-1}$;(d) An image of RM, with the color scale indicating RMs between $-$50 rad~m$^{-2}$ (red) and $+$70 rad~m$^{-2}$ (blue) as indicated to the right of the image.}
\label{figfour}
\end{figure}

\begin{figure}
\includegraphics[angle=-90, width=140mm]{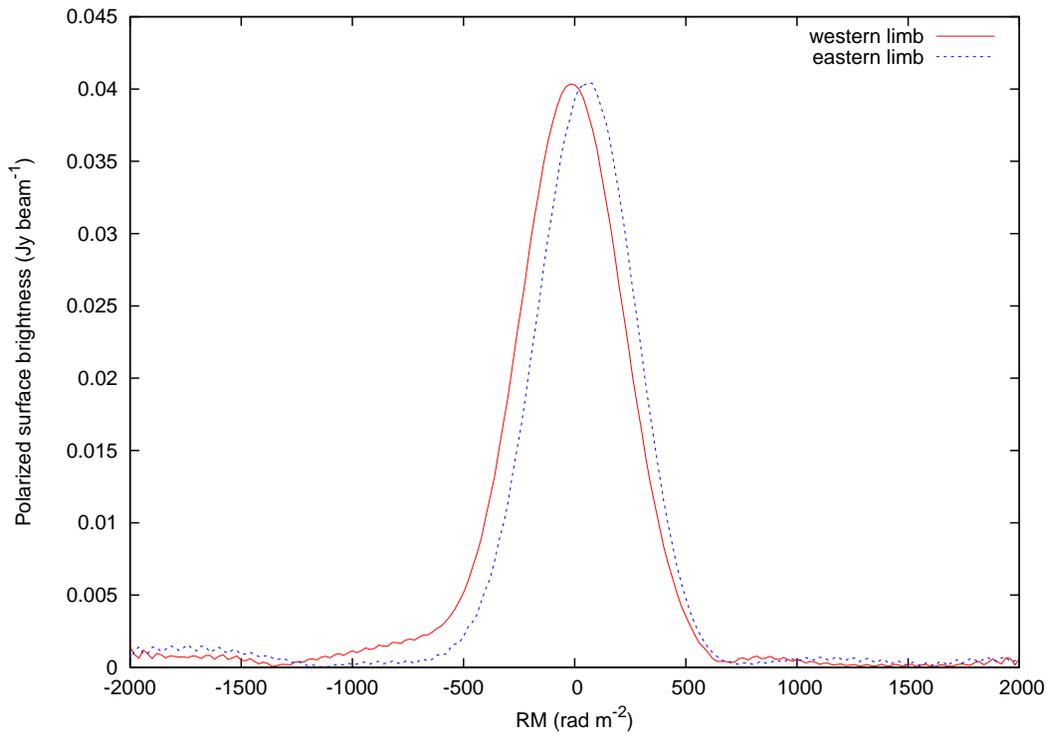}
\caption{The RM synthesis spectra of representative pixels in the eastern (blue) and western (red) limbs of SNR~\g296. Spectral deconvolution has been applied to remove the sidelobes caused by incomplete sampling in $\lambda^2$ space.}
\label{figrmspectrum}
\end{figure}

 Figure \ref{figfour} presents the images of (a) total intensity, (b) total polarized intensity (the peak $P$ as a function of RM in the RM synthesis spectrum) and (c) \& (d) peak rotation measure for SNR~\g296.
We see a clear bilateral morphology in both the total intensity and the polarized intensity images, with bright limbs of emission to the east and west of the SNR symmetry axis. Along this axis, there appears to be a complete absence of radio emission. The mean fractional polarization of \g296, found by comparing the flux density of the total polarized intensity and the Stokes-\textit{I} images at each pixel, is approximately 35$\%$. This is likely an upper limit, as our observations are not sensitive to large-scale total power emission, as discussed in \S\ref{obs}. Figure \ref{figfour} (c) shows the rotation measures as boxes, denoting the RM value averaged over five adjacent pixels. Filled boxes, representing positive RMs, dominate the eastern limb and unfilled boxes (negative RMs) domiate the western limb of the remnant.  Figure \ref{figfour} (d) shows the same information as an image. Figure~\ref{figrmspectrum} shows a typical rotation measure spectrum for a pixel in each of the eastern and western limbs of \g296. The polarized flux density is similar in each limb of the SNR, but the RM changes from positive to negative on either side of the symmetry axis.

\begin{figure}
\includegraphics[angle=-90, bb=0 0 600 600, width=105mm]{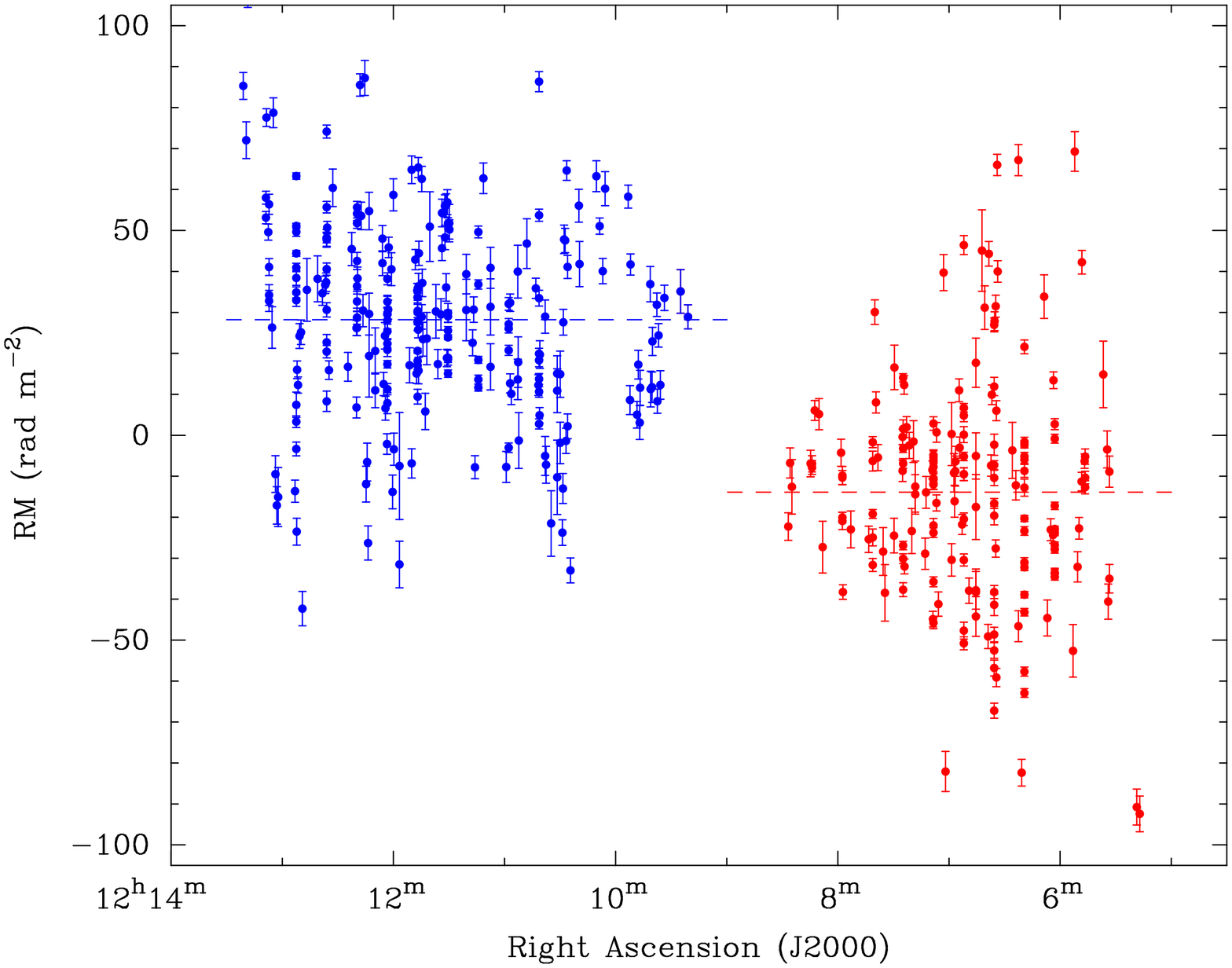}
\vfill
\includegraphics[angle=-90, bb=0 0 600 600, width=105mm]{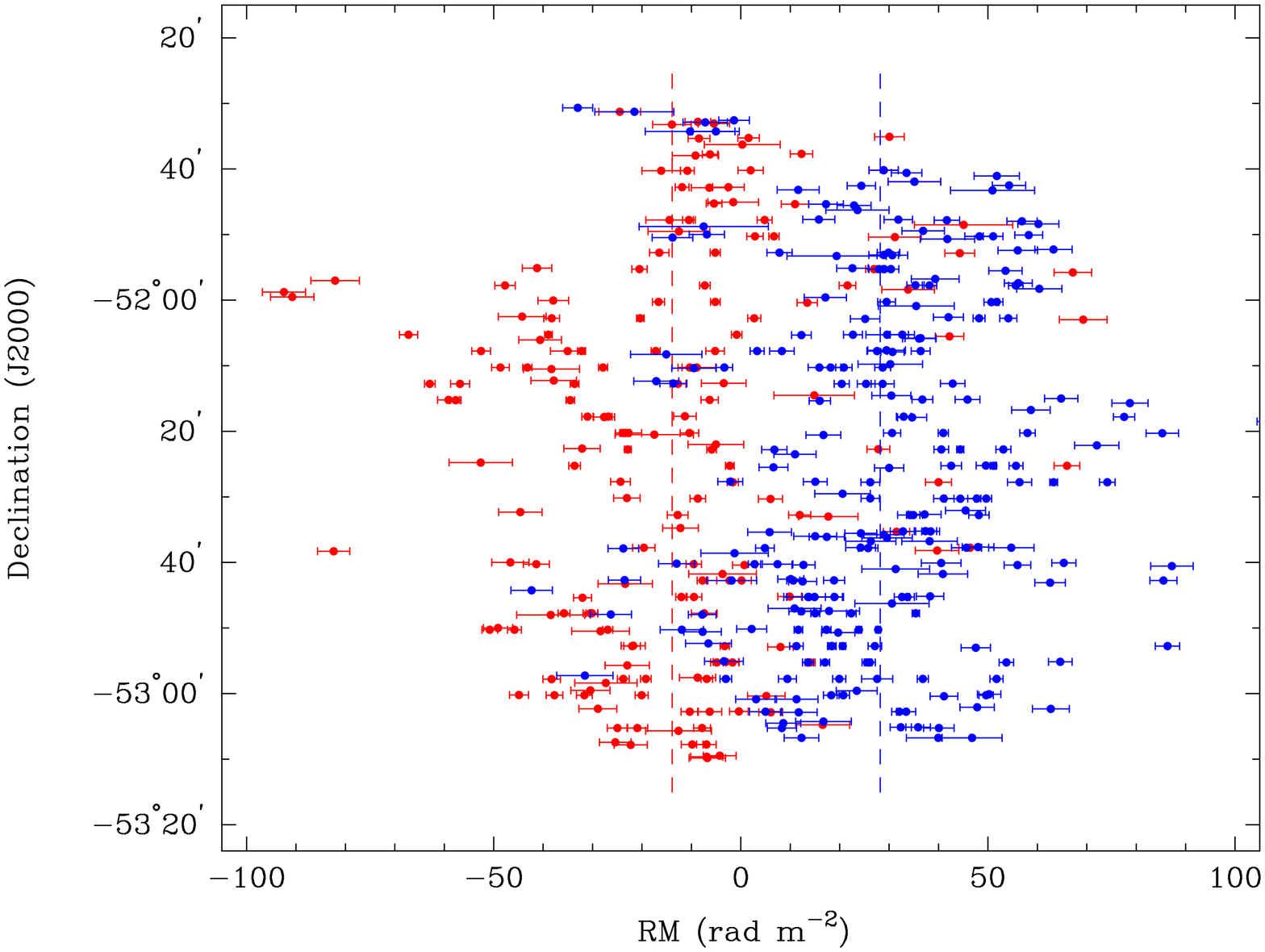}
\caption{Plots of RM against right ascension (top) and against declination (bottom) in \g296\ for the RM distribution in Figure 1 (c); red points are in the western limb and blue points are in the eastern limb. The mean RM for each limb is displayed as a dashed line.}
\label{figrmpos}
\end{figure}

In order to quantify the distributions of rotation measures in \g296\, in Figure~\ref{figrmpos} we plot RM against Right Ascension (upper panel) and against Declination (lower panel). In both plots, the mean RM in each limb is displayed as a dashed line. The upper panel of Figure \ref{figrmpos} shows a jump in RM distribution between the limbs of the SNR. The western limb has a mean RM of $-$14 $\pm$ 80 rad~m$^{-2}$ and the eastern limb has a mean of $+$28 $\pm$ 65 rad~m$^{-2}$. In contrast, in the lower panel of Figure~\ref{figrmpos} we see no such gradient or discontinuity in RM as a function of declination. This observation is crucial to our understanding of the magnetic field in the region, which we discuss further in \S\S \ref{sec_fgd}, \ref{sec_uniform} and \ref{sec_wind}.

\section{Discussion}
\label{discussion}

\subsection{Comparison with previous results}

The distribution of RM in Figure \ref{figfour} along the limbs of
\g296\ has a clear symmetry: the eastern limb shows predominantly
positive RMs, while the western limb displays negative RMs. Since the sign of the RM corresponds
to the direction of the electron-weighted average magnetic field
between the observer and the source of emission, our results provide
a high-angular resolution view of the orientation of the magnetic
field responsible for the Faraday rotation. Specifically, the average
magnetic field vector points generally toward from us on the eastern
side, and away from us on the western side. 

We first compare our results to previous studies of polarization and Faraday rotation toward SNR~\g296. \cite{wg68} first studied the magnetic fields in \g296\ using the Parkes 64-m radio telescope to map polarized radio continuum emission from the remnant at 0.6, 1.4 and 2.7 GHz. They made maps of RM across the source by measuring the best-fit linear relationship between the polarization angle and the square of the wavelength at each image pixel. The resulting RM map showed a positive range of RMs ($+$17 to $+$36 rad~m$^{-2}$) in the eastern limb with a positive gradient of RM with increasing declination. In the western limb, they found a spread of negative ($-$4 to $-$14 rad~m$^{-2}$) RMs with no declination dependence.  \cite{dm76} used a similar method to compare the polarized radio continuum emission at 2.7 and 5.0 GHz. They noted that the remnant has a moderate east-west RM gradient and that the projected magnetic field has an orientation that runs tangential to the bright edges of the SNR. They also measured an east-west gradient in rotation measure of $+$38 to $-$18 rad~m$^{-2}$ with uncertainties of $\pm$23 rad~m$^{-2}$, which agree with the values found by \cite{wg68}. \cite{mh94b} observed \g296\ using the Parkes 64-m telescope at 2.4, 4.8 and 8.4~GHz. This study surpassed the two previous polarization observations of the region, having both a higher frequency and angular resolution than measured previously. They too found strong evidence for a tangential plane of sky magnetic field. Their RM maps indicated largely negative RMs across the majority of the remnant, but with a large patch of positive RM values around the middle of the eastern limb.

 Our results agree broadly with the east-west gradient of RMs found by \cite{wg68} and \cite{dm76}, although we find no evidence for a declination dependence of RM in the eastern limb. However, our RMs are not readily compared with those derived by \cite{mh94b}; they quote a much higher range of RMs.  This may be in some part due to their cutoff level for rotation measures being 5$\%$ of the peak polarized flux at that frequency, which is a much less stringent cutoff criterion than our 10$\sigma$ in the peak polarized intensity. The upper limit on percentage polarization we derive (35$\%$) broadly agrees with that (``in excess of 30$\%$'') quoted by \cite{mh94b}. Our data were taken around 1.4~GHz, a low radio frequency where extrapolation of measured polarization angles to intrinsic values would give large ($>$ 1 radian) uncertainties. We do not provide such extrapolations here and assume the tangential distribution of intrinsic projected magnetic field vectors calculated by previous authors.

Before we interpret the polarization information gained from our observations, we need to consider whether the observed
Faraday rotation is predominantly occurring in unrelated foreground interstellar gas \citep[e.g.,][]{cmc04}, or in material that has
been swept up by the expanding SNR \citep{mldg84,kb09}.

\subsection{The RM of the Foreground ISM}
\label{sec_fgd}

Radio pulsars are typically strongly linearly polarized, and
consequently many such sources in the  ATNF Pulsar
Catalogue\footnote{On-line catalogue at {\tt
http://www.atnf.csiro.au/research/pulsar/psrcat}} \citep{mhth05}
have measured RMs. Pulsars are at a variety of distances within the
Galaxy, and their Faraday rotation is thought to have no intrinsic
component due to the pulsar itself. We can thus compare the RMs
seen against \g296\ to the RMs of pulsars in a similar direction
and at comparable distances, to assess whether the RMs seen toward
\g296\ are due to the foreground ISM.

There are two pulsars in the catalogue which sit within $10^\circ$ of
\g296, and have estimated distances between 1 and 3~kpc (bracketing the distance to the SNR): PSR~B1133--55 ($\ell = 292\fdg3$,
$b = +5\fdg9$) has RM~$ = +28\pm5$~rad~m$^{-2}$ at a distance
of $\approx2.6$~kpc \citep{qmlg95} while PSR~J1253--5820 ($\ell =
303\fdg2$, $b = +4\fdg5$) has RM~$=+18\pm1$~rad~m$^{-2}$ at a distance
$\approx2.9$~kpc \citep{njkk08}, where the distance estimates come from
the NE2001 Galactic electron density of \cite{cl02}. If we assume that the
observed RM is the projection of a magnetic field predominantly parallel
to the Galactic plane, and that the field is sufficiently uniform along
these sightlines that the magnitude of the RM scales with distance,
we can estimate from this pulsar data that the RM of the foreground ISM
between Earth and the SNR is approximately +10 to +20 rad~m$^{-2}$.

We do not find a good match between this expectation and the observed
distribution of RMs against the SNR. The RMs of the western limb are
of the opposite sign to the pulsar RMs, while those in the east are of
the correct sign, but typically have a magnitude twice as large as what
would be expected from the pulsar data.

Additional evidence against a foreground interpretation for the SNR RMs is
the relative uniformity of the sign of the RM distribution along each limb (shown in the lower panel of Figure~\ref{figrmpos}),
but the difference in the sign of RM between limbs (upper panel of Figure~\ref{figrmpos}).  If this Faraday
rotation occurs in the foreground, it requires a complete reversal of the
magnetic field on a relatively small scale in the east-west direction,
but no such reversals along a north-south axis.  This situation
must be reasonably localized, because no such reversal is seen in the
RMs of pulsars or of extragalactic sources at the same Galactic longitude
but at lower Galactic latitudes \citep{bhg+07}.  While such ``magnetic
anomalies'' have been seen previously in RM \citep[e.g.,][]{bt01},
they tend to be associated with the complicated distribution of ionized
gas and magnetic fields along the Galactic plane. At this relatively high
Galactic latitude, with no other extended sources or features anywhere in
the vicinity, it is hard to see what could cause the required foreground
RM distribution.  It would also need to be a convenient coincidence that
the RM distribution of an unrelated foreground feature has a symmetry
axis that aligns with that defined by the SNR's morphology.  We conclude
that the RM pattern seen against \g296\ is most likely intrinsic to the
source, and must therefore result from material swept by the expansion
of the SNR into its surroundings.

\subsection{The RM of a Swept-Up Uniform Ambient Medium}
\label{sec_uniform}

\cite{kb09} have pointed out that ambient gas swept up into a shell by an
expanding SNR can produce an observable RM signal, and that the resulting
systematic variations in RM as a function of position around the SNR can
be used to infer the {\em in situ}\ magnetic field geometry of the gas
into which the SNR has exploded.  However, there are two
difficulties arising if the RMs seen toward \g296\ are interpreted in
this way.

First, the observed pattern of RMs is completely contrary to
expectations. If a SNR expands into a well-ordered ambient field
with a significant component in the plane of the sky, a barrel
morphology should result as is observed here \citep{laa62b,wg68},
but there will be identical RM gradients along each
SNR limb, with the gradient aligned with the SNR's symmetry axis
\citep{kb09}. We observe the exact opposite here: there is a minimal
RM gradient along the SNR's symmetry axis, but a large gradient across it (Figure~\ref{figrmpos}). This field geometry is difficult to explain by a model in which the SNR sweeps up a uniform magnetic field.

Second, the magnitude of the RM through the shell implied by this scenario
is much larger than observed.  To illustrate this, we assume that the RM
pattern of the SNR is due to a geometric effect that is occurring equally
in both limbs (deferring a specific interpretation of this pattern until
\S\ref{sec_wind} below). We then compare the RMs of the two limbs to
separate the foreground RM contribution from that due to material in
the SNR shell: the average RM between the two limbs is approximately
$+$5 to $+$10~rad~m$^{-2}$ (which is roughly consistent with the foreground
contribution estimated in \S\ref{sec_fgd} above), while the difference
between the limbs is $\approx 40$~rad~m$^{-2}$. Thus we estimate that
the observed RM contribution of each limb is $\approx 20$~rad~m$^{-2}$,
with the average field directed in opposite directions on each side to
produce the observed RM pattern.

Assuming that the synchrotron-emitting and Faraday-rotating regions in the
swept-up shell are mixed together, the region will not be Faraday thin,
but rather will emit a continuous spread of superimposed RMs, ranging
from a foreground-only RM from material at the edge of the SNR closest
to Earth to a maximal RM value experienced by polarized signals that
propagate from the back of the SNR, through the swept-up shell, and then
through the foreground ISM.  The observed RMs and the variation between
them are both much smaller than the width of the RM transfer function
in our observations of $\sim500$~rad~m$^{-2}$. We thus cannot resolve the
source as Faraday thick, but instead will measure an average RM equal to
half the full value through the SNR \citep{gw66,sbs+98}.  Thus in relating
the observations to the electron density and magnetic field strength in
the swept-up limbs, the observed RM of $\approx 20$~rad~m$^{-2}$ implies
an RM through the SNR shell of ${\rm RM_{ISM}} \approx 40$~rad~m$^{-2}$.

To compare this to the expected value of ${\rm RM_{ISM}}$, we assume
that a SNR of radius $R$~pc has expanded into a medium of uniform
hydrogen density $n_0$~cm$^{-3}$.  If the SNR shock compresses the
material that it sweeps up by a factor $X$, then the thermal electron
density within the material shocked by the SNR (assuming 100\% ionization
of ambient gas by the SNR shock) is $n_1 = X n_0$ where the subscript 1 denotes the post-shock state.  Conservation of
mass requires that the thickness of the swept-up shell is $\Delta R =
R/3X$, while the maximum depth of a sightline through the SNR limb is
$L \approx 2(2R\Delta R)^{1/2}$ (assuming $\Delta R \ll R$). The RM
through the entire limb is then $0.81 n_1  B_1 L \cos \theta$, where
$B_1$ is the magnetic field strength in the post-shock gas, and $\theta$
is the angle of the post-shock field to the line of sight.  The maximal
RM for the swept-up material, for $\theta = 0$, is then ${\rm RM_{ISM}}
\approx1.3 X^{1/2} n_0 B_1 R$~rad~m$^{-2}$.

At a distance of 2.1~kpc, a representative radius for the SNR (measured
from the geometric center to the brightest regions of the limbs to
the east and west) is $R \approx 20$~pc.  \HI\ and X-ray observations
of the SNR imply $n_0 \sim 0.1$~cm$^{-3}$ \citep{dcg86,mlt88}, while
equipartition between relativistic particles and magnetic fields in
the emitting region \citep[e.g.,][]{pac70} implies strength in the
SNR shell of $B_1 \approx25-50$~$\mu$G.  \cite{rmk+88} argue that
SNR~\g296\ is in the adiabatic (Sedov-Taylor) phase of evolution, for
which $X=4$; the implied shell thickness $\Delta R = R/12$ is consistent
with the observed SNR morphology.  We then predict ${\rm RM_{ISM}}
\approx 130-260$~rad~m$^{-2}$, which is substantially larger than the
value ${\rm RM_{ISM}} \approx 40$~rad~m$^{-2}$ required by observations. From the above considerations, we conclude that an ambient uniform medium
cannot explain either the strength or pattern of the observed RMs.

\subsection{The RM of the Swept-Up Progenitor Wind}
\label{sec_wind}

An alternative explanation for the RMs observed in \g296\ is Faraday rotation through the magneto-ionic medium of the swept-up
stellar wind from the supernova progenitor.  At large distances
from the stellar surface, stellar winds are expected to have largely toroidal
fields; if such a field geometry is preserved after material has
been swept up by the supernova explosion, this could potentially
produce an RM pattern that was negative on one limb of the SNR and
positive on the other.  We now consider this possibility in detail.

The presence of a central neutron star in \g296\ \citep[e.g.,][]{hb84,mbc96} demonstrates that the
SNR resulted from the core-collapse of a massive star.  These stars
have powerful winds that can result in substantial mass loss well before
the supernova explosion, and which can extend for many parsecs from the
star \citep{loz92,glm96,gml96} It is thus both theoretically expected and
well-established from observations that many SNRs are propagating through
their progenitor stellar winds, with consequent implications for an SNR's
dynamics and emission properties \citep{che82b,cl89,dwa07,wps+07}. Indeed,
such a possibility has been specifically suggested for \g296\ by
\cite{ssmk92}, to explain the high degree of mirror symmetry between
the two limbs.

It is also well-established that at least some massive stars harbor strong magnetic fields \citep[e.g.][]{wade03}, and that these stars correspondingly launch magnetized winds
\citep{wd67,icb98,uo02}.
The effects on an SNR expanding into a magnetized wind were
discussed by \cite{cl94}, but the resulting RM of the swept-up wind was
not addressed. Here we provide a first simple calculation of the likely
Faraday rotation signature of this interaction; a subsequent study will
further explore this phenomenon through a detailed magnetohydrodynamical
calculation.

As noted above, an azimuthal magnetic field naturally produces
Faraday rotation that is positive on one side of the center and
negative on the other. For this field to have been produced by the
stellar wind of the progenitor, we need to assume that the
symmetry axis of the SNR corresponds to the spin-axis of the
progenitor \cite[see][]{ssmk92} and that the field pattern in the
wind is preserved when it is swept up by the SNR. 

However, we note that the simplest geometry for a magnetized stellar
wind is a split monopole, in which the magnetic field configurations
in the northern and southern hemispheres have opposite polarities,
separated by a current sheet \citep{sak85,mes03}. For such a wind, the azimuthal
component of the pre-supernova magnetic field will be directed in
opposite directions above and below the equatorial plane of the
progenitor.  The simplistic expectation is that a swept-up shell
of such material would show a quadrupolar pattern of RMs with, for
example, positive RMs to the northwest and southeast, and negative
RMs to the northeast and southwest. This is not the pattern that
we observe in Figure~\ref{figfour}.

For a magnetized swept-up wind to be a viable explanation of the
observed RMs, we thus require a higher-order field in the progenitor
wind, such as an aligned quadrupole. Such a wind, as may occur in
at least some supernova progenitors \citep{tl85}, can have an azimuthal
field of one polarity in a broad equatorial region covering both
northern and southern sides of the equatorial plane, with regions
of alternating polarity toward the poles.  Since the barrel
morphology of SNR~\g296\ is dominated by two limbs on either side
of the symmetry axis, this is a geometry that can potentially match
the observed RM pattern.  Further investigations of this wind
geometry and its interaction with a SNR shock are needed, but for
the purposes of the discussion below, we assume that this or other
simple wind patterns can produce the RMs seen in our data, and we
now consider whether the magnitudes of the RMs are consistent with
expectations from a swept-up stellar wind.

For a spherical stellar wind of mass-loss rate $\dot{M}$ and asymptotic
velocity $V_\infty$, conservation of mass implies that the number density $n(r)$ of gas at a radius $r$ from
the center of the star (where $r$ lies outside the wind acceleration region) is:
\begin{equation}
n(r) = \frac{\dot{M}}{4\pi r^2 V_\infty m_H},
\end{equation}
where $m_H$ is the mass of a hydrogen atom.

For a rotating star with a magnetic field, the (presumed ionized) stellar wind is channeled by the magnetic field lines i.e. the plasma is frozen into the field. Magnetic field lines leave the surface of the star radially and rapidly wind up into a spiral shape at radii greater than the maximum radius of co-rotation. For a star of radius $R_*$ with an equatorial rotational velocity $V_{rot}$ and assuming symmetry above and below the equatorial plane, the strength, $B_\phi(r)$, of the azimuthal component of 
the stellar wind's magnetic
field \citep{lc99} is:
\begin{equation}
B_\phi(r) = B_* \left(\frac{R_*}{r}\right)
\left(\frac{V_{rot}}{V_\infty}\right),
\end{equation}
for $r \gg R_*$, where  $B_*$ is the magnetic field strength at $r=R_*$.

We now consider the result of an SNR expanding through this wind,
so that the gas and magnetic field distributions described by these
equations are swept up into a thin shell of radius $R$ and thickness
$\Delta R$. For the purposes of the present discussion, we make the
simplifying assumption that after this material is swept up, the
shocked material within the shell has a uniform density $n_1$ and a
uniform magnetic field strength $B_1$, as for the simpler case
discussed in \S\ref{sec_uniform} above.

The total number of atoms in the wind, $N_{tot}$, swept up by the SNR  is:
\begin{equation}
N_{tot} = \int_{R_*}^{R} n(r) 4 \pi r^2 dr = \frac{\dot{M} R}{V_\infty m_H}
\end{equation}

\noindent for $r \gg$ $R_*$.

\noindent Conservation of mass requires that $N_{tot} = 4 \pi R^2 \Delta R n_1$, so
that:
\begin{equation}
n_1 = \frac{\dot{M}}{4\pi R^2 V_\infty m_H} \left( \frac{R}{\Delta R} \right)
= n(R) \left(\frac{R}{\Delta R}\right).
\label{eqn_n1}
\end{equation}

\noindent We can similarly consider the total magnetic flux,
$\Phi_{tot}$, passing through a meridional (constant $\Phi$) area element that extends in radius from $R_*$ to $R$ and spans $\Delta\psi$ in colatitude: 
\begin{equation}
\Phi_{tot} = \int_{R_*}^R B_\phi(r) r dr = B_* R_* R \Delta\psi \left(\frac{V_{rot}}{V_\infty}\right).
\end{equation}

\noindent where $\psi$ is the polar angle of the spherical co-ordinate system. 

\noindent As the material is ejected radially, the meridional area is radially compressed but the polar angle is unchanged before and after the compression.
 
Assuming conservation of flux after the SNR sweeps up this material,
we require $\Phi_{tot} = B_1 R \Delta R \Delta \psi$ and hence:
\begin{equation}
B_{\phi,1} = B_* \left(\frac{R_*}{\Delta R}\right) 
\left(\frac{V_{rot}}{V_\infty}\right) =
B_{\phi}(R) \left(\frac{R}{\Delta R}\right).
\label{eqn_b1}
\end{equation}
We are now in a position to estimate the RM produced by the swept-up wind,
\begin{equation}
{\rm RM_{wind}} = 0.81 \int \left( \frac{n_1}{{\rm cm}^{-3}} \right)
\left( \frac{B_{\phi,1}}{\mu{\rm G}} \right) 
\left(\frac{dl}{\rm pc} \right) \cos \theta ~{\rm rad~m}^{-2}
\label{eqn_rm}
\end{equation}
where the integral is along a path-length $dl$ through the shell, $\theta$
is the angle of $B_1$ to the line of sight, and where we have assumed
that the swept-up wind is fully ionized.  As in \S\ref{sec_uniform}, the
maximum path length through the limb is $L\approx 2(2R \Delta R)^{1/2}$.
Through the limbs of the SNR, a swept-up azimuthal field will be oriented
such that $\theta \approx 0$. Substituting
Equations~(\ref{eqn_n1}) and (\ref{eqn_b1}) into 
Equation~(\ref{eqn_rm}), and writing $x = V_{rot} / V_\infty$, we derive:
\begin{eqnarray}
{\rm RM_{wind}} = 3.6 \times10^{-6} \left( \frac{R}{\Delta R} \right)^{3/2}
\left( \frac{B_*}{\rm G} \right) \left( \frac{R_*}{100 R_\odot} \right)
\nonumber \\
\left( \frac{x}{0.1} \right) \left( \frac{\dot{M}}{10^{-6} M_\odot~{\rm yr}^{-1}} \right)
\left( \frac{R}{20~{\rm pc}} \right)^{-2} 
\left( \frac{V_\infty}{1000~{\rm km~s}^{-1}} \right)^{-1}~{\rm rad~m}^{-2}.
\label{eqn_rm2}
\end{eqnarray}
This equation clearly
implies that a wind from
a main-sequence or Wolf-Rayet supernova progenitor will
not produce significant Faraday rotation when it is swept up, because 
the stellar radius is small ($\le10 R_\odot$),
the mass-loss rate is low ($\dot{M} < 10^{-5}-10^{-6} M_\odot$~yr$^{-1}$) and
the wind velocity is high ($V_\infty > 1000$~km~s$^{-1}$) \citep[see][and references
therein]{cl89,dwa07,cro07}.
However, if we adopt representative values
for a red supergiant and its wind,
$B_* = 500$~G, $R_* = 300 R_\odot$, $x=0.1$, $\dot{M} = 1 \times 10^{-5} M_\odot$~yr$^{-1}$
and $V_\infty = 30$~km~s$^{-1}$ 
\citep{gt87,dor04,vvd05,vch09,van09},
we predict:
\begin{equation}
{\rm RM_{wind}} = 1.8 \left( \frac{R}{\Delta R} \right)^{3/2}~{\rm rad~m}^{-2},
\end{equation}
For a shell thickness $R / \Delta R \approx 8$ as indicated
by the radio morphology of the SNR,
this yields ${\rm RM_{wind}}  \approx 40$~rad~m$^{-2}$,
which is a factor of two larger than the observed RM, as required
(see discussion about Faraday depth effects in \S\ref{sec_uniform}).
We note that a red supergiant wind flowing
outward for $\sim 1$~Myr \citep[e.g.][]{whw02} at
$V_\infty \approx 30$~km~s$^{-1}$ will extend $\sim 30$~pc  from the
star, so that an SNR with a current radius $R\approx20$~pc would still be
propagating through this wind. 


We caution that Equation~(\ref{eqn_rm2}) has many free parameters,
and that supernova progenitors show a very wide range of properties.
Thus we cannot conclusively demonstrate or rule out that the RM seen
against the SNR matches the expected Faraday rotation from its progenitor
wind. However, having ruled out other reasonable alternatives
for the observed RM pattern in \S\S\ref{sec_fgd} \& \ref{sec_uniform},
and having shown that the RM for at least one set of reasonable wind
parameters can match the observations, we conclude that an azimuthal
magnetic field in the stellar wind of a red supergiant progenitor can
potentially explain both the sign and magnitude of the RMs seen against
SNR~\g296.

As an additional test for this scenario of a swept-up red supergiant wind, we now consider whether the barrel-shape of \g296\ is consistent with our proposed model. In order to do this, we will examine the two possible mechanisms by which a magnetized red supergiant progenitor star can produce a bilateral SNR.  These are: (i) expansion of the supernova into a magnetized stellar wind and (ii) expansion of the supernova into a stellar wind with a biconical outflow aligned with the spin axis \citep{m87}.

In the first case, the circumstellar material is distributed axisymmetrically by the action of the magnetic field \citep{sak85} and is swept up by the expanding supernova shock into a barrel-shaped SNR. Similarly, in the second case the supernova sweeps up material in all directions \emph{except} for a bipolar outflow cavity, which has been largely evacuated of material due to the high wind velocity along this axis. Under either of these schemes, the result is a remnant with two bright limbs (one at either side of the spin axis of the star) and two null points along the axis. 


These two models for the creation of barrel-shaped SNRs both involve a supernova expanding through an axisymmetric distribution of circumstellar material. A prominent example of such a system is the young barrel-shaped supernova remnant of SN1987A. This remnant is understood to have been shaped by the interaction of the supernova shock with the anisotropic circumstellar medium caused by the red and blue supergiant phases of the progenitor star \citep{dwa07a}. We argue that a supernova expanding into a magnetically-driven stellar wind is the most plausible scenario and explains the structure of \g296. This notion is supported by the findings of \cite{ssmk92}, who argued that the small-scale correlations in structure on either side of the SNR originate from changes in the outflow from the progenitor star rather than an external ISM.
 
An additional consideration is the large axial ratio of SNR~\g296. Although \cite{rmk+88} attributed the elongation of the limbs to coupling of material with the magnetic field of the ambient ISM, a model involving a SNR expanding into a magetized stellar wind can just as readily explain this phenomenon. The distribution of circumstellar material resulting from a magnetized progenitor stellar wind will result in a large fraction of material being deposited in the magnetic equatorial plane and progressively less mass distributed towards the polar regions. A spherically symmetric supernova explosion occuring within such a medium will result in a remnant with some `straightening' of the limbs, because when the shell is being swept up, the pressure experienced by the SNR shell encountering the equatorial CSM will be greater than the pressure encountered by that part encountering less dense material towards the polar regions \citep{blc96}. This picture is consistent with the small-scale structure symmetry noted by \cite{ssm+92} in \g296.

A final piece of the puzzle is the radio polarization observation of
SNR~\g296\ by \cite{mh94b}. By observing at higher radio frequencies than
we present here, they were able to infer the intrinsic orientation of
the polarization vectors in the SNR's synchrotron radiation, before this
emission is Faraday rotated in foreground thermal gas. They found that
the magnetic field orientation of the synchrotron-emitting regions has
a projection on the sky which is largely tangential to the rim of the
SNR. This can be explained in the context of a magnetized progenitor
wind if we consider a progenitor star with its spin axis tilted with respect to the plane of the sky and with an azimuthal magnetic field. Such an alignment would produce a supernova remnant with linearly polarized emission that is aligned with the sky-plane magnetic field direction tangential to the remnant limbs, as well as a line-of-sight magnetic field that reversed direction between the two separate limbs, as observed from our Faraday rotation measurements.

\section{Conclusions}
\label{conclusions}
Using rotation measure synthesis, we have made Faraday rotation maps of the supernova remnant \g296. The limbs of the remnant were found to have RMs of opposite signs and separated by approximately 40 rad~m$^{-2}$, implying that the magnetic field in the foreground to the polarized SNR emission is oriented in opposite directions on either side of the SNR's symmetry axis.  By comparing the published RMs of pulsars observed through similar sight-lines through the Galaxy, we demonstrated that the spread of RMs within the remnant is inconsistent with Faraday rotation by the foreground magneto-ionized medium. We also considered whether a swept-up ambient interstellar medium can explain the RM distribution in \g296, but found a large discrepancy between the observed rotation measures and those predicted by theory. Finally, we found that the swept-up wind of a red supergiant progenitor star could explain the observed magnitude and distribution of RMs in the remnant, as well as the linear polarization found by other authors. This would require an aligned quadropole or similar magnetic field morpology in the wind of the progenitor star.

Future observational studies should use RM synthesis to map Faraday rotation in a larger sample of Galactic supernova remnants. Focusing on remnants that lie at relatively large Galactic latitudes will minimize the effect of the Galactic foreground and enable an accurate determination of the intrinsic properties of the magnetized plasma within the remnants. If similar morphologies are found, we will be able to gain a deeper understanding about the mass-loss histories of progenitor stars prior to their supernova events. If the Faraday rotation in the remnant represents the signature of a magnetized stellar wind, then the method suggested by \cite{kb09} for studying ambient magnetic fields in the Galaxy using SNRs is not only a powerful \textit{in situ} probe of magnetic fields, but also of SNR progenitor winds. By studying a larger sample of sources we can hope to understand the physical origin of magnetic fields swept up in supernova remnants. In parallel, magnetohydrodynamical simulations of red supergiant winds should also be employed to test the validity of our model of a swept-up magnetized stellar wind in \g296.

\begin{acknowledgements}

We gratefully acknowledge Stan Owocki, Asif ud-Doula and Wouter
Vlemmings for helpful discussions on magnetic fields in massive stars
and their winds. The Australia Telescope is funded by the Commonwealth
of Australia for operation as a National Facility managed by CSIRO.
L.H-S.\ and B.M.G.\ acknowledge the support of the Australian Research
Council through grants FF0561298 and DP0986386.

\end{acknowledgements}

{\it Facilities:} \facility{ATCA}

\bibliographystyle{apj}
\bibliography{harveysmith}

\begin{thebibliography}{86}
\expandafter\ifx\csname natexlab\endcsname\relax\def\natexlab#1{#1}\fi

\bibitem[{{Bisnovatyi-Kogan} {et~al.}(1990){Bisnovatyi-Kogan}, {Lozinskaia}, \&
  {Silich}}]{bls90}
{Bisnovatyi-Kogan}, G.~S., {Lozinskaia}, T.~A., \& {Silich}, S.~A. 1990, \apss,
  166, 277

\bibitem[{{Blondin} {et~al.}(1996){Blondin}, {Lundqvist}, \&
  {Chevalier}}]{blc96}
{Blondin}, J.~M., {Lundqvist}, P., \& {Chevalier}, R.~A. 1996, \apj, 472, 257

\bibitem[{Brentjens \& de~Bruyn(2005)}]{bd05}
Brentjens, M.~A. \& de~Bruyn, A.~G. 2005, A\&A, 441, 1217

\bibitem[{{Brentjens} \& {de Bruyn}(2005)}]{bdb05}
{Brentjens}, M.~A. \& {de Bruyn}, A.~G. 2005, \aap, 441, 1217

\bibitem[{Brown {et~al.}(2007)Brown, Haverkorn, Gaensler, Taylor, Bizunok,
  McClure-Griffiths, Dickey, \& Green}]{bhg+07}
Brown, J.~C., Haverkorn, M., Gaensler, B.~M., Taylor, A.~R., Bizunok, N.~S.,
  McClure-Griffiths, N.~M., Dickey, J.~M., \& Green, A.~J. 2007, ApJ, 663, 258

\bibitem[{{Brown} \& {Taylor}(2001)}]{bt01}
{Brown}, J.~C. \& {Taylor}, A.~R. 2001, ApJ, 563, L31

\bibitem[{Burton {et~al.}(1988)Burton, Geballe, Brand, \& Webster}]{bgbw88}
Burton, M.~G., Geballe, T.~R., Brand, P. W. J.~L., \& Webster, A.~S. 1988,
  MNRAS, 231, 617

\bibitem[{{Caswell} {et~al.}(2004){Caswell}, {McClure-Griffiths}, \&
  {Cheung}}]{cmc04}
{Caswell}, J.~L., {McClure-Griffiths}, N.~M., \& {Cheung}, M.~C.~M. 2004,
  MNRAS, 352, 1405

\bibitem[{{Chevalier}(1974)}]{che74}
{Chevalier}, R.~A. 1974, \apj, 188, 501

\bibitem[{{Chevalier}(1982)}]{che82c}
---. 1982, \apjl, 259, L85

\bibitem[{Chevalier(1982)}]{che82b}
Chevalier, R.~A. 1982, ApJ, 258, 790

\bibitem[{Chevalier(1999)}]{che99}
---. 1999, ApJ, 511, 798

\bibitem[{{Chevalier} \& {Fransson}(1987)}]{cf87}
{Chevalier}, R.~A. \& {Fransson}, C. 1987, \nat, 328, 44

\bibitem[{Chevalier \& Liang(1989)}]{cl89}
Chevalier, R.~A. \& Liang, E.~P. 1989, ApJ, 344, 332

\bibitem[{Chevalier \& Luo(1994)}]{cl94}
Chevalier, R.~A. \& Luo, D. 1994, ApJ, 421, 225

\bibitem[{{Chevalier} \& {Oishi}(2003)}]{co03}
{Chevalier}, R.~A. \& {Oishi}, J. 2003, \apjl, 593, L23

\bibitem[{{Claussen} {et~al.}(1997){Claussen}, {Frail}, {Goss}, \&
  {Gaume}}]{cfg+97}
{Claussen}, M.~J., {Frail}, D.~A., {Goss}, W.~M., \& {Gaume}, R.~A. 1997, \apj,
  489, 143

\bibitem[{{Cordes} \& {Lazio}(2002)}]{cl02}
{Cordes}, J.~M. \& {Lazio}, T.~J.~W. 2002, preprint (arXiv:astro-ph/0207156)

\bibitem[{Crowther(2007)}]{cro07}
Crowther, P.~A. 2007, Ann. Rev. Astr. Ap., 45, 177

\bibitem[{Dickel \& Milne(1976)}]{dm76}
Dickel, J.~R. \& Milne, D.~K. 1976, Aust. J. Phys., 29, 435

\bibitem[{Dorsch(2004)}]{dor04}
Dorsch, S. B.~F. 2004, A\&A, 423, 1101

\bibitem[{Dubner {et~al.}(1986)Dubner, Colomb, \& Giacani}]{dcg86}
Dubner, G.~M., Colomb, F.~R., \& Giacani, E.~B. 1986, AJ, 91, 343

\bibitem[{{Dwarkadas}(2005)}]{dwa05}
{Dwarkadas}, V.~V. 2005, \apj, 630, 892

\bibitem[{Dwarkadas(2007)}]{dwa07}
Dwarkadas, V.~V. 2007, ApJ, 667, 226

\bibitem[{{Dwarkadas}(2007)}]{dwa07a}
{Dwarkadas}, V.~V. 2007, in American Institute of Physics Conference Series,
  Vol. 937, Supernova 1987A: 20 Years After: Supernovae and Gamma-Ray Bursters,
  ed. {S.~Immler, K.~Weiler, \& R.~McCray}, 120--124

\bibitem[{{Foster} {et~al.}(2004){Foster}, {Routledge}, \& {Kothes}}]{frk04}
{Foster}, T., {Routledge}, D., \& {Kothes}, R. 2004, \aap, 417, 79

\bibitem[{{Frail} {et~al.}(1994){Frail}, {Goss}, \& {Slysh}}]{fgs94}
{Frail}, D.~A., {Goss}, W.~M., \& {Slysh}, V.~I. 1994, \apjl, 424, L111

\bibitem[{{Franco} {et~al.}(1991){Franco}, {Tenorio-Tagle}, {Bodenheimer}, \&
  {Rozyczka}}]{ftb+91}
{Franco}, J., {Tenorio-Tagle}, G., {Bodenheimer}, P., \& {Rozyczka}, M. 1991,
  \pasp, 103, 803

\bibitem[{{Fukui} {et~al.}(2003){Fukui}, {Moriguchi}, {Tamura}, {Yamamoto},
  {Tawara}, {Mizuno}, {Onishi}, {Mizuno}, {Uchiyama}, {Hiraga}, {Takahashi},
  {Yamashita}, \& {Ikeuchi}}]{fmt+03}
{Fukui}, Y., {Moriguchi}, Y., {Tamura}, K., {Yamamoto}, H., {Tawara}, Y.,
  {Mizuno}, N., {Onishi}, T., {Mizuno}, A., {Uchiyama}, Y., {Hiraga}, J.,
  {Takahashi}, T., {Yamashita}, K., \& {Ikeuchi}, S. 2003, \pasj, 55, L61

\bibitem[{Gaensler(1998)}]{gae98}
Gaensler, B.~M. 1998, ApJ, 493, 781

\bibitem[{{Gaensler} {et~al.}(1998){Gaensler}, {Manchester}, \&
  {Green}}]{gmg98}
{Gaensler}, B.~M., {Manchester}, R.~N., \& {Green}, A.~J. 1998, \mnras, 296,
  813

\bibitem[{Gaensler {et~al.}(1997)Gaensler, Manchester, Staveley-Smith,
  Tzioumis, Reynolds, \& Kesteven}]{gms+97}
Gaensler, B.~M., Manchester, R.~N., Staveley-Smith, L., Tzioumis, A.~K.,
  Reynolds, J.~E., \& Kesteven, M.~J. 1997, ApJ, 479, 845

\bibitem[{Garc\'{i}a-Segura {et~al.}(1996{\natexlab{a}})Garc\'{i}a-Segura,
  Langer, \& Mac~Low}]{glm96}
Garc\'{i}a-Segura, G., Langer, N., \& Mac~Low, M.-M. 1996{\natexlab{a}}, A\&A,
  316, 133

\bibitem[{Garc\'{i}a-Segura {et~al.}(1996{\natexlab{b}})Garc\'{i}a-Segura,
  Mac~Low, \& Langer}]{gml96}
Garc\'{i}a-Segura, G., Mac~Low, M.-M., \& Langer, N. 1996{\natexlab{b}}, A\&A,
  305, 229

\bibitem[{Gardner \& Whiteoak(1966)}]{gw66}
Gardner, F.~F. \& Whiteoak, J.~B. 1966, Ann. Rev. Astr. Ap., 4, 245

\bibitem[{{Giacani} {et~al.}(2000){Giacani}, {Dubner}, {Green}, {Goss}, \&
  {Gaensler}}]{gdg+00}
{Giacani}, E.~B., {Dubner}, G.~M., {Green}, A.~J., {Goss}, W.~M., \&
  {Gaensler}, B.~M. 2000, \aj, 119, 281

\bibitem[{Gray \& Toner(1987)}]{gt87}
Gray, D.~F. \& Toner, C.~G. 1987, ApJ, 322, 360

\bibitem[{Green {et~al.}(1997)Green, Frail, Goss, \& Otrupcek}]{gfgo97}
Green, A.~J., Frail, D.~A., Goss, W.~M., \& Otrupcek, R. 1997, AJ, 114, 2058

\bibitem[{{Gvaramadze}(2006)}]{gva06}
{Gvaramadze}, V.~V. 2006, \aap, 454, 239

\bibitem[{{Heald} {et~al.}(2009){Heald}, {Braun}, \& {Edmonds}}]{hbe09}
{Heald}, G., {Braun}, R., \& {Edmonds}, R. 2009, \aap, 503, 409

\bibitem[{Helfand \& Becker(1984)}]{hb84}
Helfand, D.~J. \& Becker, R.~H. 1984, Nature, 307, 215

\bibitem[{Ignace {et~al.}(1998)Ignace, Cassinelli, \& Bjorkman}]{icb98}
Ignace, R., Cassinelli, J.~P., \& Bjorkman, J.~E. 1998, ApJ, 505, 910

\bibitem[{Koo \& Heiles(1995)}]{kh95}
Koo, B. \& Heiles, C. 1995, ApJ, 442, 679

\bibitem[{Kothes \& Brown(2009)}]{kb09}
Kothes, R. \& Brown, J. 2009, in Cosmic Magnetic Fields: From Planets, to Stars
  and Galaxies (IAU Symposium 259), ed. K.~G. Strassmeier, A.~G. Kosovichev, \&
  J.~E. Beckman (Cambridge: Cambridge University Press), 75--80

\bibitem[{Koyama {et~al.}(1995)Koyama, Petre, Gotthelf, Hwang, Matsura, Ozaki,
  \& Holt}]{kpg+95}
Koyama, K., Petre, R., Gotthelf, E.~V., Hwang, U., Matsura, M., Ozaki, M., \&
  Holt, S.~S. 1995, Nature, 378, 255

\bibitem[{{Kulsrud} {et~al.}(1965){Kulsrud}, {Bernstein}, {Krusdal}, {Fanucci},
  \& {Ness}}]{kbk+65}
{Kulsrud}, R.~M., {Bernstein}, I.~B., {Krusdal}, M., {Fanucci}, J., \& {Ness},
  N. 1965, \apj, 142, 491

\bibitem[{{Lamers} \& {Cassinelli}(1999)}]{lc99}
{Lamers}, H.~J.~G.~L.~M. \& {Cassinelli}, J.~P. 1999, {Introduction to Stellar
  Winds} (Cambridge: Cambridge University Press)

\bibitem[{{Landecker} {et~al.}(1999){Landecker}, {Routledge}, {Reynolds},
  {Smegal}, {Borkowski}, \& {Seward}}]{lrr+99}
{Landecker}, T.~L., {Routledge}, D., {Reynolds}, S.~P., {Smegal}, R.~J.,
  {Borkowski}, K.~J., \& {Seward}, F.~D. 1999, \apj, 527, 866

\bibitem[{Lozinskaya(1992)}]{loz92}
Lozinskaya, T.~A. 1992, Supernovae and Stellar Wind in the Interstellar Medium
  (New York: American Institute of Physics)

\bibitem[{{Manchester}(1987)}]{m87}
{Manchester}, R.~N. 1987, \aap, 171, 205

\bibitem[{{Manchester} {et~al.}(2005){Manchester}, {Hobbs}, {Teoh}, \&
  {Hobbs}}]{mhth05}
{Manchester}, R.~N., {Hobbs}, G.~B., {Teoh}, A., \& {Hobbs}, M. 2005, AJ, 129,
  1993

\bibitem[{{Matsui} {et~al.}(1984){Matsui}, {Long}, {Dickel}, \&
  {Greisen}}]{mldg84}
{Matsui}, Y., {Long}, K.~S., {Dickel}, J.~R., \& {Greisen}, E.~W. 1984, ApJ,
  287, 295

\bibitem[{{Matsui} {et~al.}(1988){Matsui}, {Long}, \& {Tuohy}}]{mlt88}
{Matsui}, Y., {Long}, K.~S., \& {Tuohy}, I.~R. 1988, ApJ, 329, 838

\bibitem[{Mereghetti {et~al.}(1996)Mereghetti, Bignami, \& Caraveo}]{mbc96}
Mereghetti, S., Bignami, G.~F., \& Caraveo, P.~A. 1996, ApJ, 464, 842

\bibitem[{Mestel(2003)}]{mes03}
Mestel, L. 2003, Stellar Magnetism (Oxford: Oxford University Press)

\bibitem[{Milne \& Haynes(1994)}]{mh94b}
Milne, D.~K. \& Haynes, R.~F. 1994, MNRAS, 270, 106

\bibitem[{Noutsos {et~al.}(2008)Noutsos, Johnston, Kramer, \&
  Karastergiou}]{njkk08}
Noutsos, A., Johnston, S., Kramer, M., \& Karastergiou, A. 2008, MNRAS, 386,
  1881

\bibitem[{Pacholczyk(1970)}]{pac70}
Pacholczyk, A.~G. 1970, Radio Astrophysics (San Francisco: Freeman)

\bibitem[{{Pineault} {et~al.}(1993){Pineault}, {Landecker}, {Madore}, \&
  {Gaumont-Guay}}]{plm+93}
{Pineault}, S., {Landecker}, T.~L., {Madore}, B., \& {Gaumont-Guay}, S. 1993,
  \aj, 105, 1060

\bibitem[{Qiao {et~al.}(1995)Qiao, Manchester, Lyne, \& Gould}]{qmlg95}
Qiao, G.~J., Manchester, R.~N., Lyne, A.~G., \& Gould, D.~M. 1995, MNRAS, 274,
  572

\bibitem[{Roger {et~al.}(1988)Roger, Milne, Kesteven, Wellington, \&
  Haynes}]{rmk+88}
Roger, R.~S., Milne, D.~K., Kesteven, M.~J., Wellington, K.~J., \& Haynes,
  R.~F. 1988, ApJ, 332, 940

\bibitem[{Sakurai(1985)}]{sak85}
Sakurai, T. 1985, A\&A, 152, 121

\bibitem[{{Sault} {et~al.}(1999){Sault}, {Bock}, \& {Duncan}}]{sbd99}
{Sault}, R.~J., {Bock}, D.~C.-J., \& {Duncan}, A.~R. 1999, \aaps, 139, 387

\bibitem[{Sault \& Killeen(2004)}]{sk03}
Sault, R.~J. \& Killeen, N. E.~B. 2004, The Miriad User's Guide (Sydney:
  Australia Telescope National Facility),
  (http://www.atnf.csiro.au/computing/software/miriad/)

\bibitem[{Sokoloff {et~al.}(1998)Sokoloff, Bykov, Shukurov, Berkhuijsen, Beck,
  \& Poezd}]{sbs+98}
Sokoloff, D.~D., Bykov, A.~A., Shukurov, A., Berkhuijsen, E.~M., Beck, R., \&
  Poezd, A.~D. 1998, MNRAS, 299, 189

\bibitem[{{Storey} {et~al.}(1992){Storey}, {Staveley-Smith}, {Manchester}, \&
  {Kesteven}}]{ssm+92}
{Storey}, M.~C., {Staveley-Smith}, L., {Manchester}, R.~N., \& {Kesteven},
  M.~J. 1992, \aap, 265, 752

\bibitem[{Storey {et~al.}(1992)Storey, Staveley-Smith, Manchester, \&
  Kesteven}]{ssmk92}
Storey, M.~C., Staveley-Smith, L., Manchester, R.~N., \& Kesteven, M.~J. 1992,
  A\&A, 265, 752

\bibitem[{{Tanimori} {et~al.}(1998){Tanimori}, {Hayami}, {Kamei}, {Dazeley},
  {Edwards}, {Gunji}, {Hara}, {Hara}, {Holder}, {Kawachi}, {Kifune}, {Kita},
  {Konishi}, {Masaike}, {Matsubara}, {Matsuoka}, {Mizumoto}, {Mori}, {Moriya},
  {Muraishi}, {Muraki}, {Naito}, {Nishijima}, {Oda}, {Ogio}, {Patterson},
  {Roberts}, {Rowell}, {Sakurazawa}, {Sako}, {Sato}, {Susukita}, {Suzuki},
  {Suzuki}, {Tamura}, {Thornton}, {Yanagita}, {Yoshida}, \&
  {Yoshikoshi}}]{thk+98}
{Tanimori}, T., {Hayami}, Y., {Kamei}, S., {Dazeley}, S.~A., {Edwards}, P.~G.,
  {Gunji}, S., {Hara}, S., {Hara}, T., {Holder}, J., {Kawachi}, A., {Kifune},
  T., {Kita}, R., {Konishi}, T., {Masaike}, A., {Matsubara}, Y., {Matsuoka},
  T., {Mizumoto}, Y., {Mori}, M., {Moriya}, M., {Muraishi}, H., {Muraki}, Y.,
  {Naito}, T., {Nishijima}, K., {Oda}, S., {Ogio}, S., {Patterson}, J.~R.,
  {Roberts}, M.~D., {Rowell}, G.~P., {Sakurazawa}, K., {Sako}, T., {Sato}, Y.,
  {Susukita}, R., {Suzuki}, A., {Suzuki}, R., {Tamura}, T., {Thornton}, G.~J.,
  {Yanagita}, S., {Yoshida}, T., \& {Yoshikoshi}, T. 1998, \apjl, 497, L25

\bibitem[{Thompson \& Landstreet(1985)}]{tl85}
Thompson, I.~B. \& Landstreet, J.~D. 1985, ApJ, 289, L9

\bibitem[{{Tian} \& {Leahy}(2008)}]{tl08}
{Tian}, W.~W. \& {Leahy}, D.~A. 2008, \mnras, 391, L54

\bibitem[{{ud-Doula} \& {Owocki}(2002)}]{uo02}
{ud-Doula}, A. \& {Owocki}, S.~P. 2002, ApJ, 576, 413

\bibitem[{{ud-Doula} {et~al.}(2008){ud-Doula}, {Owocki}, \& {Townsend}}]{uot08}
{ud-Doula}, A., {Owocki}, S.~P., \& {Townsend}, R.~H.~D. 2008, \mnras, 385, 97

\bibitem[{van Belle {et~al.}(2009)van Belle, J., \& Hart}]{vch09}
van Belle, G.~T., J., C.-E.~M., \& Hart, A. 2009, MNRAS, 394, 1925

\bibitem[{van~der Laan(1962{\natexlab{a}})}]{laa62a}
van~der Laan, H. 1962{\natexlab{a}}, MNRAS, 124, 125

\bibitem[{van~der Laan(1962{\natexlab{b}})}]{laa62b}
---. 1962{\natexlab{b}}, MNRAS, 124, 179

\bibitem[{van Loon(2009)}]{van09}
van Loon, J.~T. 2009, in Hot and Cool: Bridging Gaps in Massive Star Evolution,
  ed. C.~Leitherer, P.~D. Bennett, P.~W. Morris, \& J.~T. van Loon (San
  Francisco: Astronomical Society of the Pacific), in press (arXiv:0906.485)

\bibitem[{{Vel{\'a}zquez} {et~al.}(2002){Vel{\'a}zquez}, {Dubner}, {Goss}, \&
  {Green}}]{vdg+02}
{Vel{\'a}zquez}, P.~F., {Dubner}, G.~M., {Goss}, W.~M., \& {Green}, A.~J. 2002,
  \aj, 124, 2145

\bibitem[{{Vink} {et~al.}(1998){Vink}, {Bloemen}, {Kaastra}, \&
  {Bleeker}}]{vkb98}
{Vink}, J., {Bloemen}, H., {Kaastra}, J.~S., \& {Bleeker}, J.~A.~M. 1998, \aap,
  339, 201

\bibitem[{Vlemmings {et~al.}(2005)Vlemmings, van Langevelde, \&
  Diamond}]{vvd05}
Vlemmings, W. H.~T., van Langevelde, H.~J., \& Diamond, P.~J. 2005, Mem. della
  Soc. Ast. It., 76, 462

\bibitem[{{Wade}(2003)}]{wade03}
{Wade}, G.~A. 2003, in Astronomical Society of the Pacific Conference Series,
  Vol. 305, Astronomical Society of the Pacific Conference Series, ed.
  {L.~A.~Balona, H.~F.~Henrichs, \& R.~Medupe} (San Francisco: Astronomical
  Society of the Pacific), 16

\bibitem[{Weber \& Davis(1967)}]{wd67}
Weber, E.~J. \& Davis, Jr., L. 1967, ApJ, 148, 217

\bibitem[{Weiler {et~al.}(????)Weiler, Panagia, Sramek, Van~Dyk, Williams,
  Stockdale, \& Kelley}]{wps+07}
Weiler, K.~W., Panagia, N., Sramek, R.~A., Van~Dyk, S.~D., Williams, C.~L.,
  Stockdale, C.~J., \& Kelley, M.~T. ????, 258--263

\bibitem[{Whiteoak \& Gardner(1968)}]{wg68}
Whiteoak, J.~B. \& Gardner, F.~F. 1968, ApJ, 154, 807

\bibitem[{Woosley {et~al.}(2002)Woosley, Heger, \& Weaver}]{whw02}
Woosley, S.~E., Heger, A., \& Weaver, T.~A. 2002, Rev. Mod. Phys., 74, 1015

\bibitem[{{Zavlin} {et~al.}(2004){Zavlin}, {Pavlov}, \& {Sanwal}}]{zps04}
{Zavlin}, V.~E., {Pavlov}, G.~G., \& {Sanwal}, D. 2004, ApJ, 606, 444

\bibitem[{{Zavlin} {et~al.}(2000){Zavlin}, {Pavlov}, {Sanwal}, \& {Tr{\"
  u}mper}}]{zpst00}
{Zavlin}, V.~E., {Pavlov}, G.~G., {Sanwal}, D., \& {Tr{\" u}mper}, J. 2000,
  ApJ, 540, L25

\end{thebibliography}

\end{document}